\documentstyle[epsf]{elsart}
\begin{document}

\def\spose#1{\hbox to 0pt{#1\hss}}
\def\ltapprox{\mathrel{\spose{\lower 3pt\hbox{$\mathchar"218$}}
 \raise 2.0pt\hbox{$\mathchar"13C$}}}
\def\gtapprox{\mathrel{\spose{\lower 3pt\hbox{$\mathchar"218$}}
 \raise 2.0pt\hbox{$\mathchar"13E$}}}
\def\inapprox{\mathrel{\spose{\lower 3pt\hbox{$\mathchar"218$}}
 \raise 2.0pt\hbox{$\mathchar"232$}}}

\def\dirac{{\bf \rm D}\!\!\!\!/\,}
\def\wilson{{\bf \rm W}}
\def\ham{{\bf \rm H}}
\def\gauge{{\bf \rm G}}
\def\mbham{{\cal H}}
\def\bmat{{\bf \rm B}}
\def\cmat{{\bf \rm C}}

\begin{frontmatter}

\begin{flushright}
{\normalsize FSU-SCRI-98-3}\\
\end{flushright}

\title{ 
The hermitian Wilson--Dirac operator in smooth SU(2) instanton backgrounds
}
\author{Robert G. Edwards, Urs M. Heller and Rajamani Narayanan}
\address{SCRI, The Florida State University, Tallahassee,
FL 32306-4130, USA}

\begin{abstract}
We study the spectral flow of the hermitian Wilson--Dirac operator
$\ham(m)$ as a function of $m$  in
smooth SU(2) instanton backgrounds on the lattice. For a single
instanton background 
with Dirichlet boundary conditions on $\ham(m)$, we find a level
crossing in the spectral flow of $\ham(m)$, and we find the shape of the 
crossing mode at the crossing point
to be in good agreement with the zero mode associated with
the single instanton background. 
With anti-periodic boundary conditions on $\ham(m)$, we find that
the instanton background in the singular gauge has the correct
spectral flow but the one in regular gauge does not.
We also investigate the spectral flows of two instanton and 
instanton--anti-instanton backgrounds.
\end{abstract}

\end{frontmatter}

{\bf PACS \#:}  11.15.Ha.\hfill\break

\section{Introduction}

The spectral flow
of the hermitian Wilson--Dirac operator $\ham(m)=\gamma_5\wilson(-m)$
where $\wilson(m)$ is the conventional Wilson--Dirac
operator
with bare mass $m$ 
carries information about the topological content of the background
gauge field~\cite{over1,over2} and also carries information about
chiral symmetry breaking~\cite{chiral}.
Level crossings can occur in the flow for
$m>0$ and the index of the chiral Dirac operator is equal to the
net level crossings, in the sense that fermion number is violated
by an amount equal to the index. Low lying eigenvalues of $\ham(m)$
whether they cross or not can build up a finite density
of eigenvalues at zero in the thermodynamic limit and lead to
a non-zero chiral condensate.
The spectral flow enables one to
measure the distribution of the index of the chiral Dirac operator
in lattice gauge theories. This has been done for pure
SU(2) gauge theory at one value of the lattice coupling~\cite{index}.
In the continuum, the index of the chiral Dirac operator is related
to the topology of the gauge field background.
Assuming this carries over to the lattice,
the index of the chiral Dirac operator on the lattice with SU(2) gauge
field in the fundamental representation 
is equal to the topological charge of the gauge field background.
The results in~\cite{index} provide some support to the above
assumption since the distribution of the index obtained there at a fixed
gauge coupling is in
agreement with the distribution of the topological charge obtained in
~\cite{cooling} at the same lattice coupling.
In this paper we consider smooth gauge field configurations on the
lattice which contain topological objects, with the aim of providing
further support to the above assumption. Evidence that SU(2) instantons
on the lattice result in level crossings 
was first observed in Ref.~\cite{over2}. Low lying eigenvalues of
staggered fermion operator have also been used to extract the topological
content of background gauge field 
configurations~\cite{Vink,Pugh,Laursen}. 
Low lying eigenvalues of $\ham(m)$ in smooth single instanton backgrounds
have also been investigated in
Ref.~\cite{Smith} with no reference to level crossing.

In this paper we present a 
systematic study of smooth instanton backgrounds. 
We study the spectral flow of $\ham(m)$ and focus on level
crossings to obtain information about the topological charge of
the background gauge field. Since we study smooth instanton
backgrounds with one or two instantons, 
we see isolated level crossings and no concentration of eigenvalues
near zero. This is not the case in a gauge field background
generated from a path integral measure on the lattice~\cite{chiral}.
We study the effect
of changing the size and 
the location of the instanton; namely, we consider
instantons centered on a lattice site and 
in the middle of a hypercubic cell. We study the effect of the boundary
condition of $\ham(m)$ on the spectral flow, and the effect of
the gauge choice of the instanton background. 
In instanton liquid models~\cite{liquid}
one builds up a background containing
instantons and anti-instantons by summing up gauge field configurations
of an appropriate number of instantons and anti-instantons with
each in the singular gauge. We study the spectral flow
of two instantons and an instanton--anti-instanton background
for several separations of the two
objects.

\section{The hermitian Wilson-Dirac operator} 

The hermitian Wilson-Dirac operator
$\ham(m)$ in the chiral basis is
\begin{equation}
\ham(m) = \pmatrix{\bmat - m & \cmat \cr \cmat^\dagger & -\bmat + m};
\label{eq:hamil}
\end{equation}
where
\begin{eqnarray}
\cmat_{i\alpha,j\beta}(n,n^\prime) &=& {1\over 2}
\sum_\mu \sigma_\mu^{\alpha\beta} \bigl[
U^{ij}_\mu(n)\delta_{n^\prime,n+\hat\mu} - (U^\dagger_\mu)^{ij}(n^\prime)
\delta_{n,n^\prime+\hat\mu} 
\bigr] \\
\label{eq:cmat}
\bmat_{i\alpha,j\beta}(n,n^\prime) &=& {1\over 2}\delta_{\alpha,\beta}
\sum_\mu \bigl[ 2\delta_{ij}\delta_{nn^\prime}-
U^{ij}_\mu(n)\delta_{n^\prime,n+\hat\mu} - (U^\dagger_\mu)^{ij}(n^\prime)
\delta_{n,n^\prime+\hat\mu}\bigr]
\quad
\label{eq:bmat}
\end{eqnarray}
\begin{equation}
\sigma_1=\pmatrix{0 & 1\cr 1 & 0 \cr}; \ \ 
\sigma_2=\pmatrix{0 & -i\cr i & 0 \cr}; \ \ 
\sigma_3=\pmatrix{1 & 0\cr 0 & -1 \cr}; \ \ 
\sigma_4=\pmatrix{i & 0\cr 0 & i \cr} \ ,
\label{eq:pauli}
\end{equation}
and where $n$ and $n'$ label the lattice sites, $\alpha$ and
$\beta$ are two component spinor indices, and $i$ and $j$ 
are three component color indices. 
We are interested in the eigenvalue problem
\begin{equation}
\ham(m) \phi_k(m) = \lambda_k (m) \phi_k(m)
\label{eq:eigen}
\end{equation}
as a function of $m$ on a finite lattice.
In particular we are interested in level crossings in the spectral
flow of $\ham(m)$, namely points in $m$ where $\lambda_k(m)=0$ for some
$k$. All level crossings have to occur in the region $0\le m \le 8$.
Furthermore, any level crossing in the region $0\le m\le 4$ has an opposite
level crossing in the region $4\le m \le 8$. This is due to the following
two facts. 
\begin{itemize}
\item $\ham(m)$ cannot have any zero eigenvalues for $m<0$.
\item The spectrum of $\ham(m)$ and $-\ham(8-m)$ are the same for
an arbitrary gauge field background.
\end{itemize}
The first one follows from the positive definiteness of $\bmat$~\cite{over1}.
To see the second one, we note that
\begin{equation}
- \ham(8-m,U) = \ham(m,-U) = \gauge^\dagger \ham(m,U) \gauge
\label{eq:symm}
\end{equation}
where
\begin{equation}
\gauge_{ia,jb}(n,n^\prime) = \delta_{ab}\delta_{n,n^\prime} g^{ij}(n);
\ \ \ \ 
g^{ij}(n) = \cases{ \delta_{ij} & if $\sum_\mu n_\mu$ is odd \cr
-\delta_{ij} & if $\sum_\mu n_\mu$ is even \cr }
\label{eq:unitary}
\end{equation}
is a unitary matrix.
Eqn. (\ref{eq:symm}) implies that the spectrum of $-\ham(8-m)$
and $\ham(m)$ are identical for an arbitrary gauge field background.

\section{Instanton backgrounds on the lattice}

We will consider gauge field backgrounds that contain instantons.
The SU(2) link elements on the lattice obtained from 
an exact evaluation of the path ordered integral of the
continuum instanton take the following form~\cite{Laursen}. 
In the regular gauge,
\begin{equation}
U^{\rm reg}_\mu(n) = \exp \Bigl[ia_\mu(n)\cdot\sigma \vartheta_\mu(n;\rho)\Bigr]
\label{eq:Inst_r}
\end{equation}
\begin{equation}
\vartheta_\mu(n;\rho) = {1\over \sqrt{ \rho^2 + \sum_{\nu\ne\mu} (n_\nu-c_\nu)^2}}
\tan^{-1} { \sqrt{ \rho^2 + \sum_{\nu\ne\mu} (n_\nu-c_\nu)^2} \over
\rho^2 + \sum_{\nu} (n_\nu-c_\nu)^2 + (n_\nu-c_\nu) }
\label{eq:phi}
\end{equation}
\begin{eqnarray}
a_1(n) & = & (n_4-c_4, n_3-c_3, -n_2+c_2),\cr
a_2(n) & = & (-n_3+c_3, n_4-c_4, n_1-c_1),\cr
a_3(n) & = & (n_2-c_2, -n_1+c_1, n_4-c_4),\cr
a_4(n) & = & (-n_1+c_1, -n_2+c_2, -n_3+c_3)
\label{eq:a's}
\end{eqnarray}
and in the singular gauge,
\begin{equation}
U^{\rm sing}_\mu(n) = \exp \Bigl[ib_\mu(n)\cdot\sigma 
\bigr(\vartheta_\mu(n;0)-\vartheta_\mu(n;\rho)\bigl)\Bigr]
\label{eq:Inst_s}
\end{equation}
\begin{eqnarray}
b_1(n) & = & (-n_4+c_4, n_3-c_3, -n_2+c_2),\cr
b_2(n) & = & (-n_3+c_3, -n_4+c_4, n_1-c_1),\cr
b_3(n) & = & (n_2-c_2, -n_1+c_1, -n_4+c_4),\cr
b_4(n) & = & (n_1-c_1, n_2-c_2, n_3-c_3)
\label{eq:b's}
\end{eqnarray}
$U^{\rm sing}_\mu(n)$ and $U^{\rm reg}_\mu(n)$ are related by the
gauge transformation $g(n)=(n_4+i\sigma_i n_i)/|n|$. 
Here $c$ denotes the center of the instanton, and $\rho$ is the size of
the instanton measured in lattice units. 
In the regular gauge the non-trivial winding is on the sphere at infinity;
however, in the singular gauge the non-trivial winding is
localized at $c$. Instantons in regular gauge can have their center anywhere
while instantons in singular gauge cannot have their centers on the lattice
site. This is due to the singularity present in $\vartheta(n;0)$ at $n=0$.
The anti-instanton is obtained by a parity transformation
on the instanton background. 
We will construct two instanton (instanton--anti-instanton) backgrounds  
by simply multiplying the gauge field background for a single
instanton with the gauge field background for another instanton 
(an anti-instanton) in the spirit of the dilute gas approximation.
For this we will start with the single instantons in the singular
gauge since the windings are localized at the center in this gauge and
are expected to add when we put instantons together.
We will study the spectral flow
for two different boundary
conditions on $\phi_k(m)$, namely, Dirichlet and anti-periodic. 
Dirichlet boundary conditions are motivated by the fact that the
eigenfuctions are expected to vanish far away from the points where
the instantons are localized. Anti-periodic boundary conditions will
enable us to study the effect of forcing the geometry of a torus on
a non-trivial gauge field background. We use anti-periodic boundary
conditions as opposed to periodic boundary conditions
to avoid any spurious level crossings
from almost zero momentum modes. 
We use the Lanczos algorithm~\cite{Lanczos}
 to obtain the low lying eigenvalues of
$\ham(m)$. 
We will use the Ritz functional~\cite{ritz} in the neighborhood of
level crossings to
obtain information about the eigenvectors associated with
the levels that cross. The Ritz functional has two advantages over
the Lanczos algorithm. The eigenvectors are known in addition to
the eigenvalues, and the Ritz functional properly counts any
degeneracies in the spectrum. 

\section {Spectral flow in backgrounds with a single instanton}

\begin{figure}
\epsfxsize=5.0in
\centerline{\epsffile{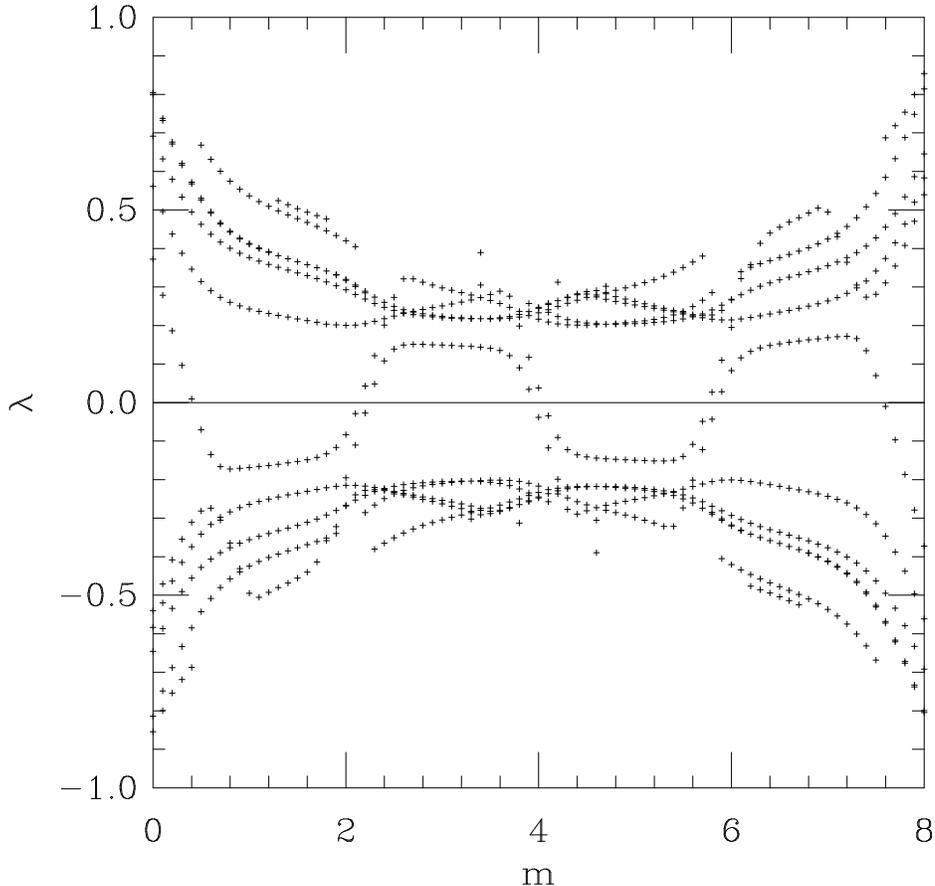}}
\caption{
Spectral flow of $\ham(m)$ for $0\le m \le 8$ for a single
instanton on an $8^4$ lattice with Dirichlet boundary conditions
for the Wilson fermions. The instanton has a size of $\rho=2.0$
and is centered at $c_\mu=4.5$. 
}
\label{fig:spec_2.0_8_o}
\end{figure}

\begin{figure}
\epsfxsize=5.0in
\centerline{\epsffile{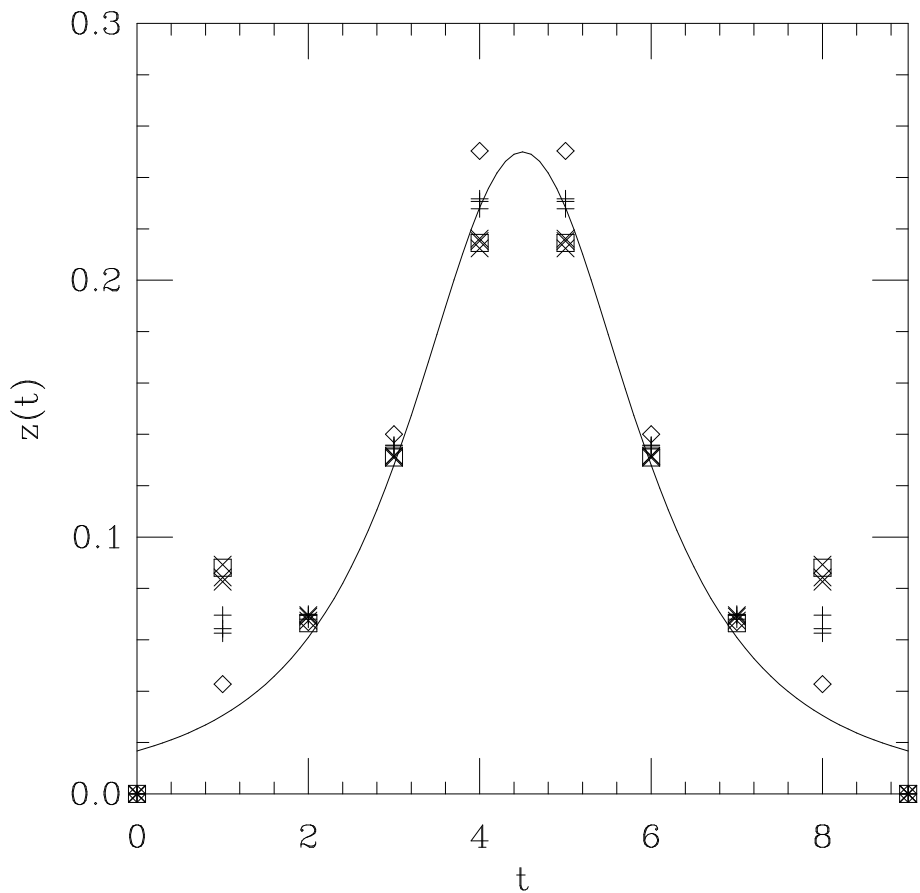}}
\caption{
The modes associated with the level crossings in
Fig.~\ref{fig:spec_2.0_8_o}. 
The diamond symbol corresponds to the crossing at $m=0.411$.
The square symbol corresponds to the crossing at $m=2.141$. The three plus
symbols correspond to the three degenerate modes crossing at $m=2.234$.
The three cross
symbols correspond to the three degenerate modes crossing at $m=3.949$.
The solid line is the zero mode for $\rho=2.0$ in the continuum.
}
\label{fig:zero_2.0_8_o}
\end{figure}

We start by considering the spectral flow of $\ham(m)$ in a single
instanton background with Dirichlet boundary conditions for $\ham(m)$.
In the continuum the chiral Dirac operator has a single zero mode
in this background~\cite{Hooft}.
If we choose $\rho >> 1$ and consider the spectral flow on a infinite
lattice, we expect to find one level crossing close to $m=0$. Since
the free Wilson-Dirac operator effectively describes massless particles
also at $m=2,4,6,8$ we expect four levels to cross near $m=2$, six
levels to cross near $m=4$ (with three a little before $m=4$ and
three a little after $m=4$), four levels to cross near $m=6$ and
finally one crossing near $m=8$. The flow with these crossings will
obey the symmetry given by~(\ref{eq:symm}). At the crossing points,
we have a zero eigenvalue of $\ham(m)$ and the associate mode should
have the appropriate chirality ($ \pm 1$)
and the shape given by the zero
mode~\cite{Hooft}. From~(\ref{eq:hamil}) and~(\ref{eq:eigen})
it follows that
\begin{equation}
{d\lambda_k(m)\over dm} = -\phi^\dagger_k(m) \gamma_5 \phi_k(m)
\label{eq:chiral}
\end{equation}
and therefore the negative of the slope of the flow at the crossing 
point gives the chirality of the zero mode.

\begin{figure}
\epsfxsize=5.0in
\centerline{\epsffile{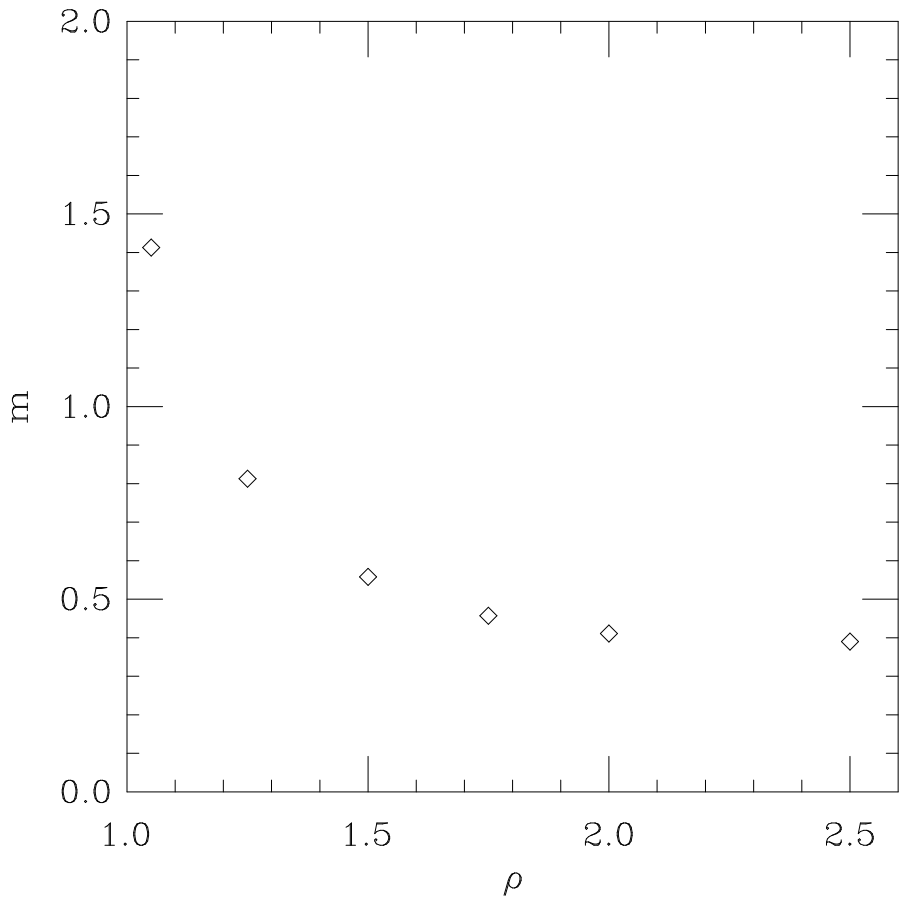}}
\caption{
The crossing point $m$ as a function of $\rho$ for a single instanton
centered at $c_\mu=4.5$ on an $8^4$ lattice with Dirichlet boundary
conditions imposed on $\ham(m)$.
}
\label{fig:rho_vs_m}
\end{figure}
To verify the above statements, we considered a $\rho=2$ instanton
on an $8^4$ lattice with $c_\mu=4.5$.\footnote{ We label the sites
of the lattice from $1$ to $8$. Therefore, the instanton is positioned
in the center of the lattice.} The flow in the region $0\le m\le 8$ 
obtained using the Lanczos algorithm~\cite{Lanczos} is
shown in Fig.~\ref{fig:spec_2.0_8_o}. We first note that the flow
obeys the symmetry given by~(\ref{eq:symm}). Using the 
Ritz functional~\cite{ritz}, we studied the modes that cross below $m=4$.
We found that the first downward crossing 
occurred at $m=0.411$. This crossing was singly degenerate and the
mode had a chirality
of $0.829$. From there, we found an upward crossing at $m=2.141$. This
crossing was also singly degenerate and the mode had a chirality of $-0.707$.
Next, we found an upward crossing at $m=2.234$. This crossing was three-fold
degenerate and all the modes had chiralities of $-0.773$. 
Finally, we found a downward crossing at $m=3.949$. This crossing was also
three-fold degenerate and all the modes had chiralities of $0.715$. 
To verify that these modes are indeed close to the continuum zero modes,
we plot $z(t)=\sum_{\vec n} \phi^\dagger_k (\vec n, t)\phi_k (\vec n, t)$
as a function of $t$ in Fig.~\ref{fig:zero_2.0_8_o} and compare it
with the zero mode in the continuum given by~\cite{Hooft} 
\begin{equation}
z(t;c,\rho)= {1\over \Bigl[ 2\rho \bigl(
1 + ({(t-c_4)\over\rho})^2 \bigr)^{3/2} \Bigr ] } 
\label{eq:Hooft_mode}
\end{equation}
The agreement is excellent. The deviations near the boundary are due to
the Dirichlet boundary conditions on the finite lattice imposed at
$t=0$ and $t=9$. 

\begin{figure}
\epsfxsize=5.0in
\centerline{\epsffile{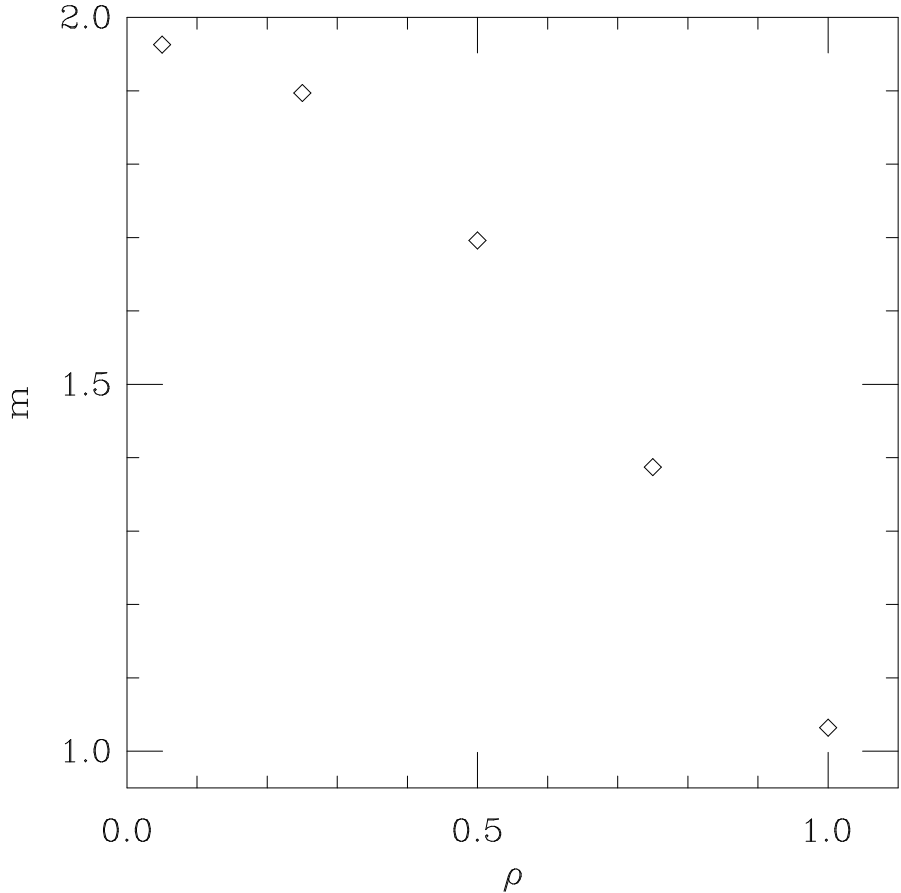}}
\caption{
The crossing point $m$ as a function of $\rho$ for a single instanton
centered at $c_\mu=5$ on an $8^4$ lattice with Dirichlet boundary
conditions imposed on $\ham(m)$.
}
\label{fig:rho_vs_m_small}
\end{figure}
The agreement between the continuum zero modes
and the modes associated with the level crossing is expected to improve
as $\rho << L \rightarrow \infty$ where $L$ is the linear extent of
the lattice. At the crossing point, we obtain the following condition
from~(\ref{eq:hamil}) and~(\ref{eq:eigen}):
\begin{equation}
u^\dagger \bmat u + v^\dagger \bmat v = m; \ \ \ \ 
\phi=\pmatrix {u\cr v\cr}
\label{eq:cross}
\end{equation}
For larger $\rho$, $\phi$ is smoother and
we expect the crossing to occur at a smaller $m$.
Fig.~\ref{fig:rho_vs_m} provides evidence for this statement.
We did not observe any level crossing for $\rho \ltapprox 1.0$ instantons
implying that instantons centered in the middle of the hypercube
``fall through the lattice'' if their size is less than about one lattice spacing.
As $\rho\rightarrow 1$ from above the crossing point in $m$ gets closer
to $m=2$. 

\begin{figure}
\epsfxsize=5.0in
\centerline{\epsffile{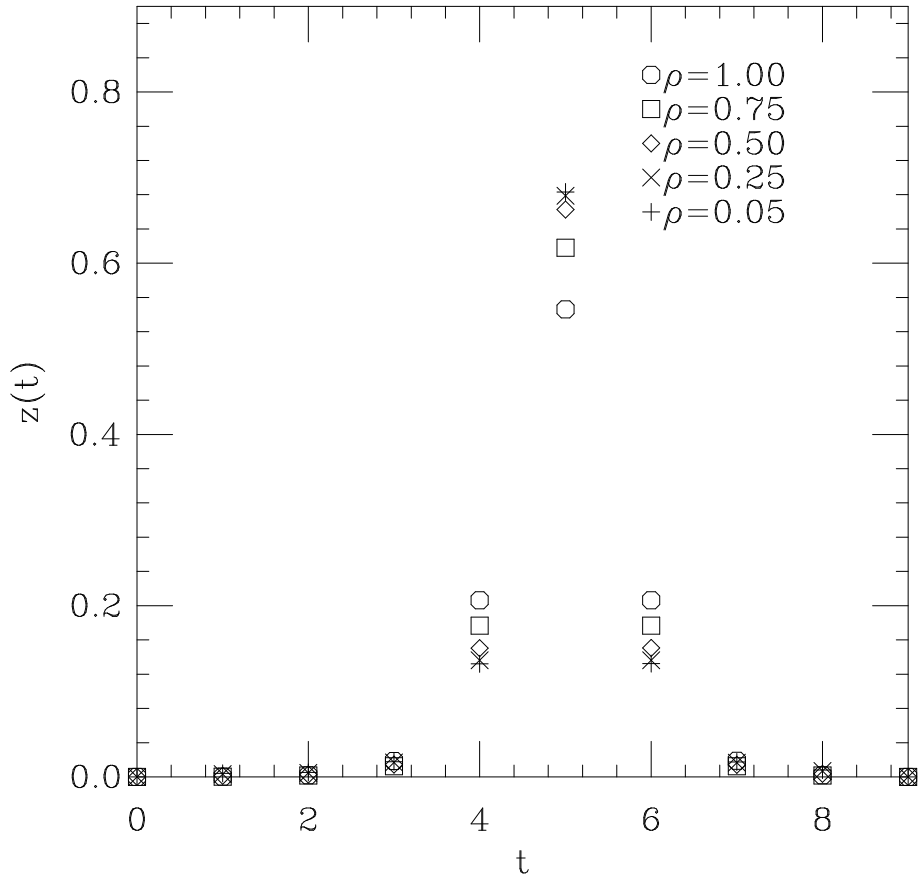}}
\caption{
The modes associated with the points in Fig.~\ref{fig:rho_vs_m_small}.
}
\label{fig:zero_vs_m}
\end{figure}
To further investigate the fate of instantons smaller than a lattice spacing
we considered small instantons centered on a lattice site as opposed
to the middle of a hypercube. In Fig.~\ref{fig:rho_vs_m_small}
we plot the crossing point as a function of $\rho$ for $\rho \le 1$ with
the instantons centered at $c_\mu=5$ on an $8^4$ lattice. We found
a level crossing even for an instanton with $\rho=0.05$, and the trend
in Fig.~\ref{fig:rho_vs_m_small} is consistent with the statement that
instantons with smaller $\rho$ result in a level crossing at a larger
$m$. 
However, site centered instantons of a given size cross for a smaller
value of $m$ compared to instantons centered in the
middle of the hypercube and having the same size. 
{\sl Instantons
do not fall through the lattice if the center of the instanton lies on
a lattice site}. 
However, the mode associated with the level crossing cannot reproduce
the continuum mode correctly. 
From~(\ref{eq:Hooft_mode}) we see that $z(t=c_4;c,\rho)=1\over{2\rho}$
and this diverges as $\rho$ approaches zero. The lattice mode
cannot reproduce this singularity. $z(t)$ for the modes associated with
the points in Fig.~\ref{fig:rho_vs_m_small} is plotted in
Fig.~\ref{fig:zero_vs_m}. Clearly, the value at $t=5$ does not grow
like $1/\rho$. In fact all these modes roughly look the same and
are consistent with a mode of size of about one. The modes at the crossing
point for a site centered instanton and a hypercube centered instanton
look alike for $\rho > 1$. 

\begin{figure}
\epsfxsize=5.0in
\centerline{\epsffile{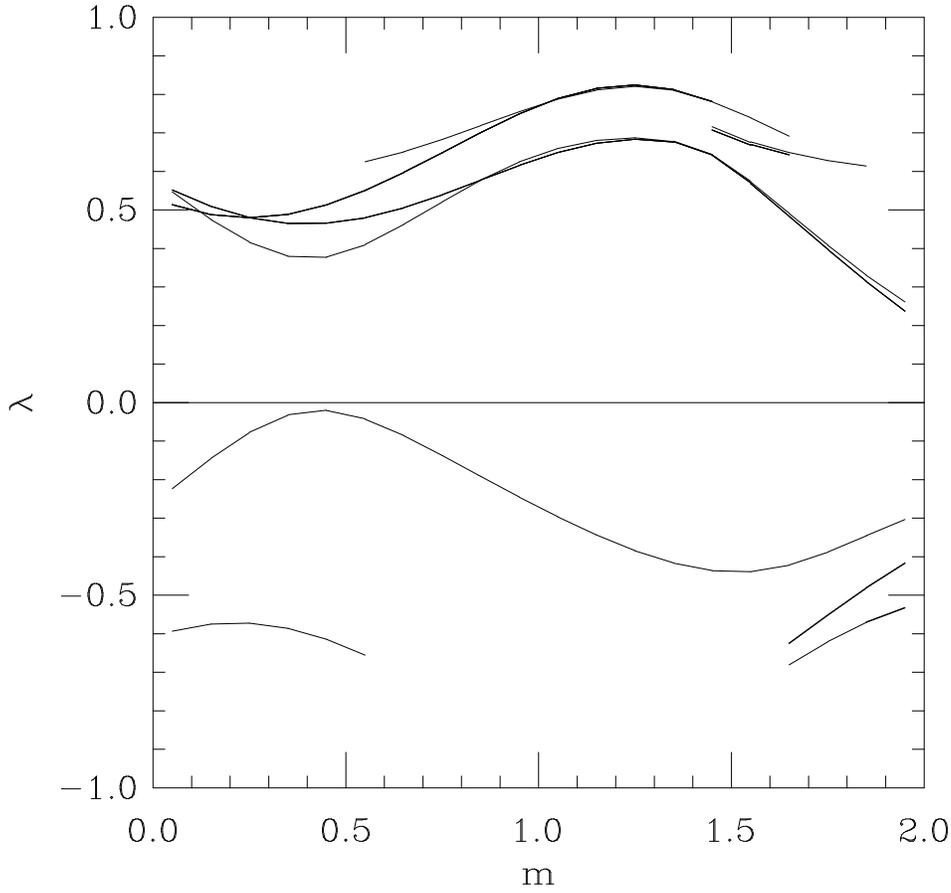}}
\caption{
Spectral flow of $\ham(m)$ for $0\le m \le 2$ for a single
instanton on an $8^4$ lattice with anti-periodic boundary conditions
for the Wilson fermions. The instanton with a size of $\rho=1.5$
and centered at $c_\mu=4.5$ is in the regular gauge. 
}
\label{fig:spec_1.5_8_p_r}
\end{figure}
\begin{figure}
\epsfxsize=5.0in
\centerline{\epsffile{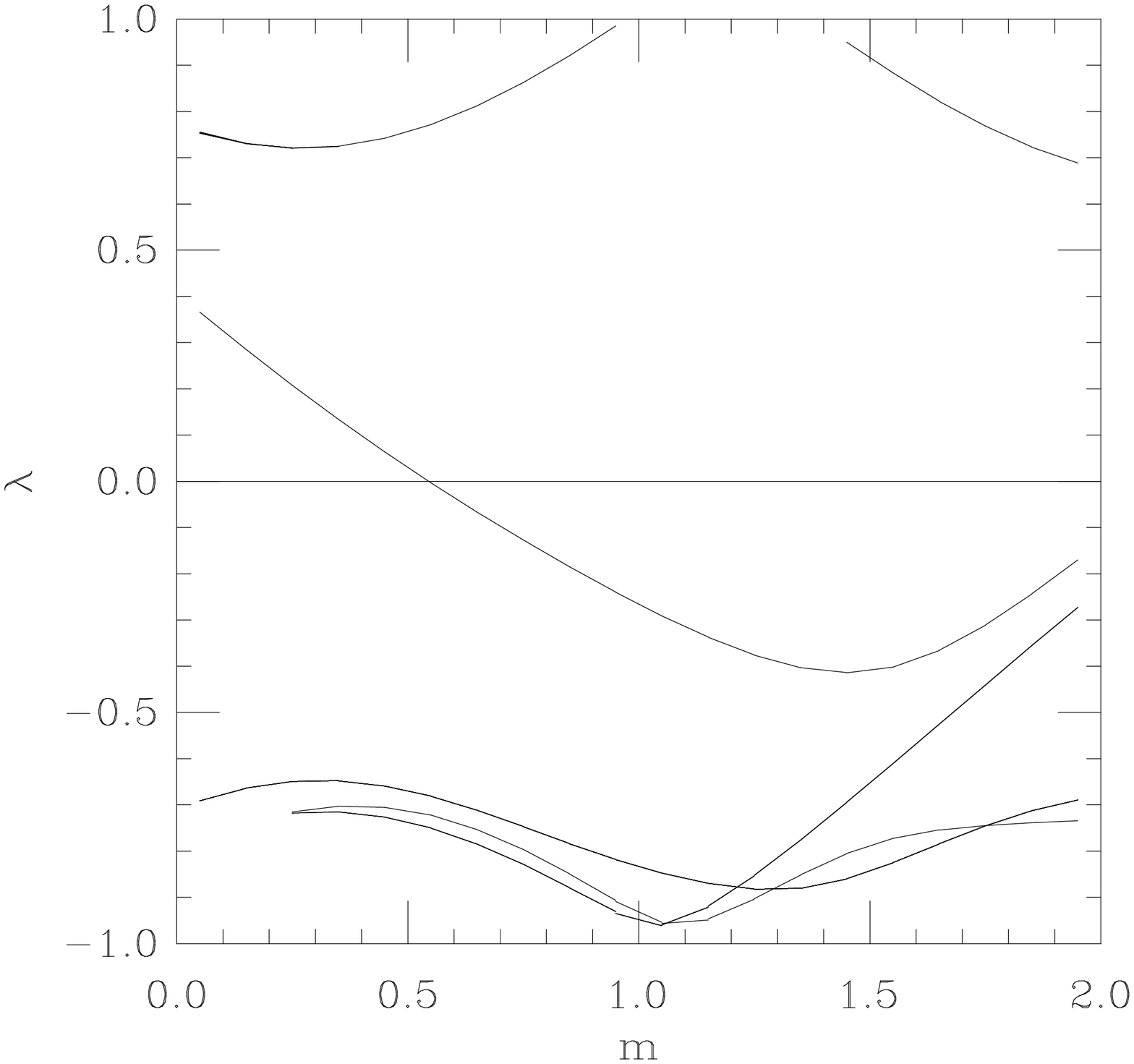}}
\caption{
Spectral flow of $\ham(m)$ for $0\le m \le 2$ for a single
instanton on an $8^4$ lattice with anti-periodic boundary conditions
for the Wilson fermions. The instanton with a  size of $\rho=1.5$
and centered at $c_\mu=4.5$ is in the singular gauge. 
}
\label{fig:spec_1.5_8_p_s}
\end{figure}
\begin{figure}
\epsfxsize=5.0in
\centerline{\epsffile{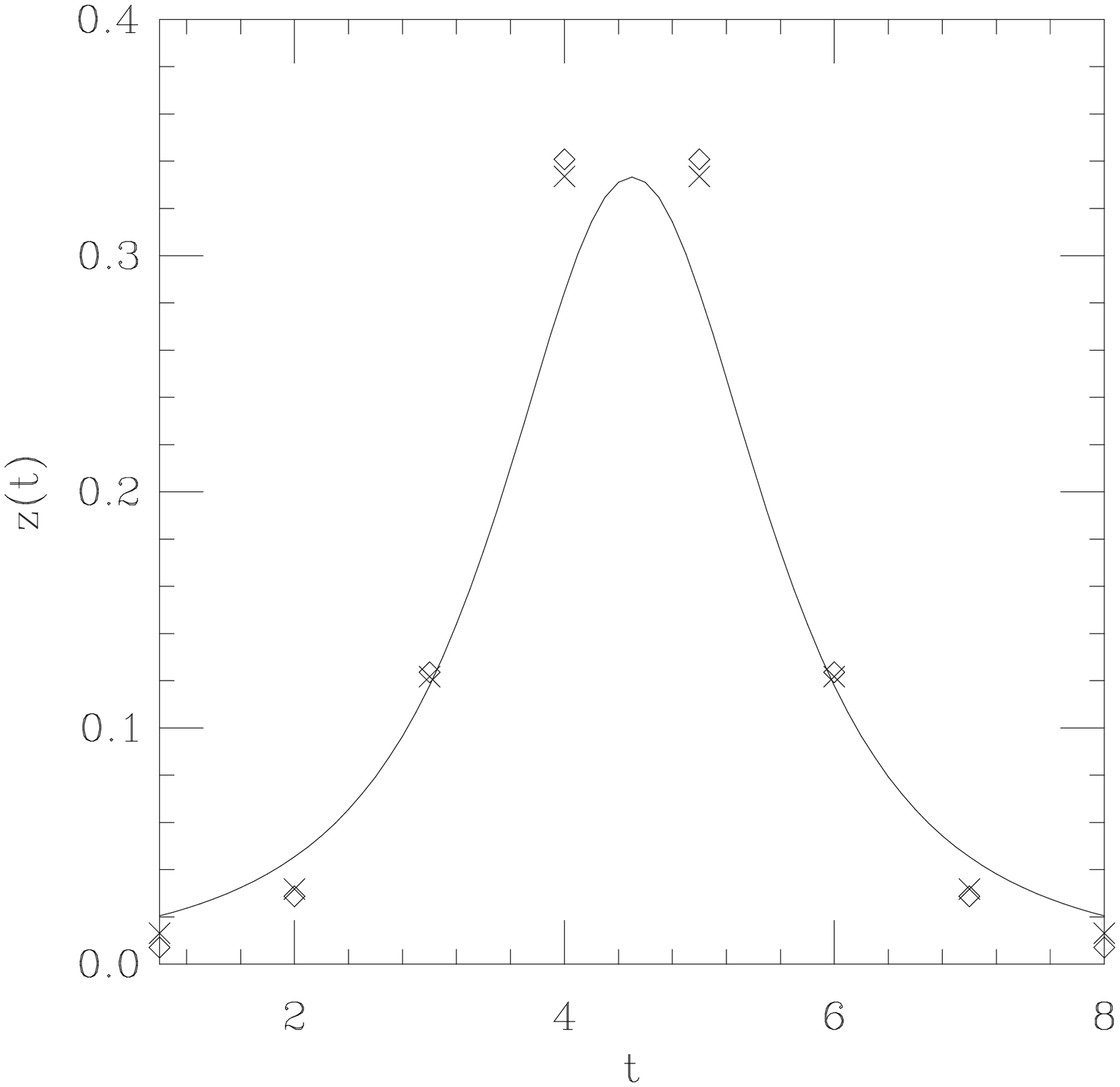}}
\caption{
The mode associated with the level crossing in
Fig.~\ref{fig:spec_1.5_8_p_s} is shown with the diamond symbol.
The corresponding mode with Dirichlet boundary conditions is shown
with the cross symbol and the solid line is the continuum
zero mode
with $\rho=1.5$.
}
\label{fig:zero_1.5_8_p_s}
\end{figure}

The instanton field in the regular gauge and the singular gauge
given by~(\ref{eq:Inst_r}) and (\ref{eq:Inst_s}), respectively,
are proper descriptions on a manifold with a spherical geometry.
The gauge potential is equivalent to zero at the origin of the sphere and on the
spherical surface at infinity. The two gauge potentials are
related by a non-trivial gauge transformation with a 
unit winding number. In the regular gauge the winding is at the
spherical surface at infinity, and in the singular gauge the
winding is localized to the point at the origin of the sphere.
In the study of the spectral flow of $\ham(m)$ with Dirichlet boundary
conditions on a finite lattice the gauge choice does not matter.
However, the gauge choice is relevant in a 
study of the spectral flow with anti-periodic boundary
conditions. Anti-periodic boundary
conditions force a toroidal geometry on a configuration that is
well defined only on a spherical geometry. Even if the size of the instanton
is much smaller than the lattice size,
we expect a
radical change in the spectral flow with anti-periodic boundary
conditions. Since the action of the gauge field configuration is
quite small, periodic boundary conditions cause spurious level
crossings due to almost zero momentum modes and this is avoided
by choosing anti-periodic boundary conditions.~\footnote{
We studied the spectral flow of $\ham(m)$ in a single instanton background
with periodic boundary conditions. We found several spurious crossings
very close to $m=0$ and the modes associated with these crossings were
very close to a constant mode. Spurious low lying eigenvalues were
also observed in the spectrum of the staggered fermion operator with
periodic boundary conditions~\cite{Laursen}.}
The spectral flow
for a $\rho=1.5$ instanton in the regular gauge
on an $8^4$ lattice with anti-periodic boundary conditions
is shown in Fig.~\ref{fig:spec_1.5_8_p_r}.
The flow for the same instanton in singular gauge is
shown in Fig.~\ref{fig:spec_1.5_8_p_s}. Fig.~\ref{fig:spec_1.5_8_p_r}
shows no crossing indicating that forcing anti-periodic boundary conditions
in regular gauge results in loss of the topology. 
However, Fig.~\ref{fig:spec_1.5_8_p_s} shows one level crossing downward
and this is consistent with the topology of a single instanton.
The mode associated with this crossing is plotted in
Fig.~\ref{fig:zero_1.5_8_p_s}. In the same figure we compare this
mode with the corresponding mode for the same instanton obtained
from the spectral flow with Dirichlet boundary conditions. The two
modes essentially lie on top of one another and both of them are
a proper characterization of the continuum zero mode as shown
in Fig.~\ref{fig:zero_1.5_8_p_s}. This shows that the topology is not lost
by forcing a toroidal
geometry if the winding is localized to the point at the center of the
instanton. This is consistent with the conclusions obtained from
a study of the staggered fermion operator~\cite{Laursen}. 
By studying the spectral flow for various instanton sizes we found that
there is no level crossing if the size is less than about one lattice
spacing.

\begin{figure}
\epsfxsize=5.0in
\centerline{\epsffile{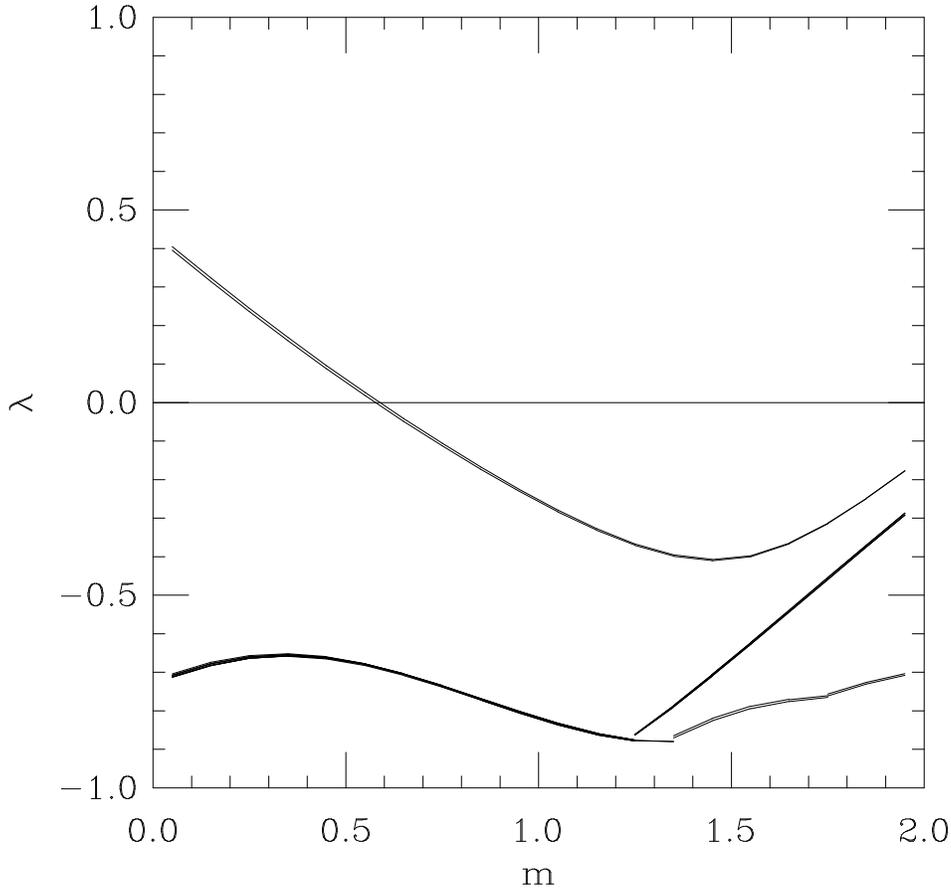}}
\caption{
Spectral flow of $\ham(m)$ for $0\le m \le 2$ for a two
instanton configuration
on an $8^4$ lattice with anti-periodic boundary conditions
for the Wilson fermions. Both instantons have a size of $\rho=1.5$
and are separated by four lattice spacings in every direction. 
}
\label{fig:spec_II}
\end{figure}
\begin{figure}
\epsfxsize=5.0in
\centerline{\epsffile{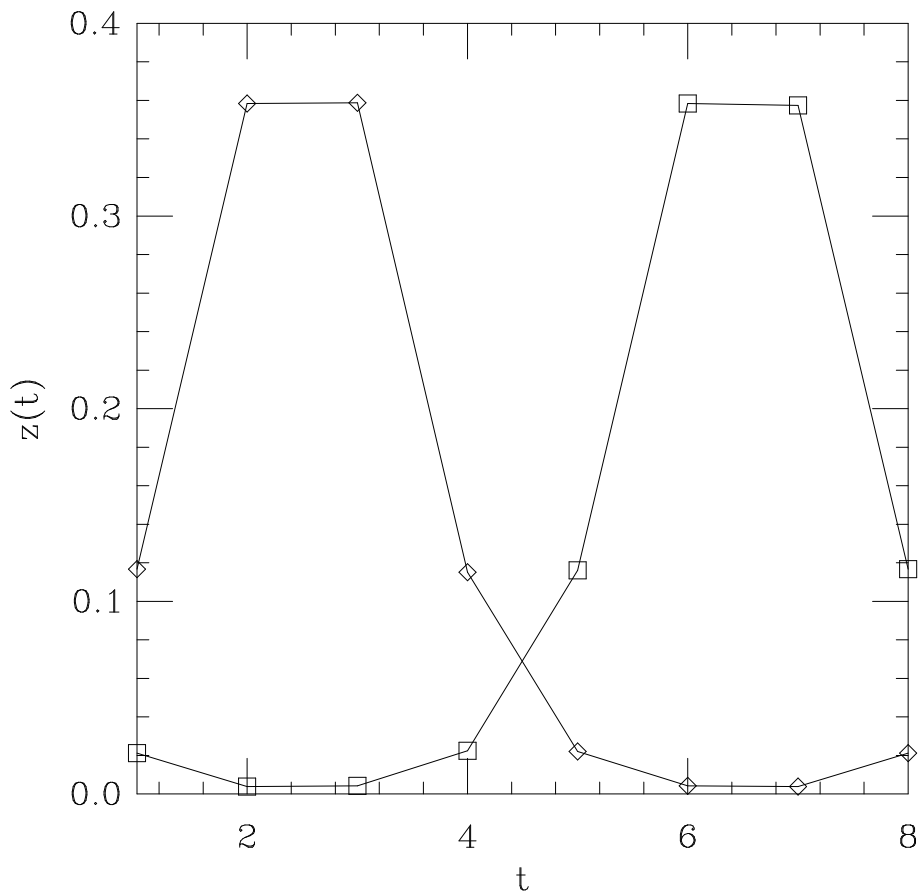}}
\caption{
The modes associated with the level crossings in
Fig.~\ref{fig:spec_II}.
}
\label{fig:zero_II}
\end{figure}
\begin{figure}
\epsfxsize=5.0in
\centerline{\epsffile{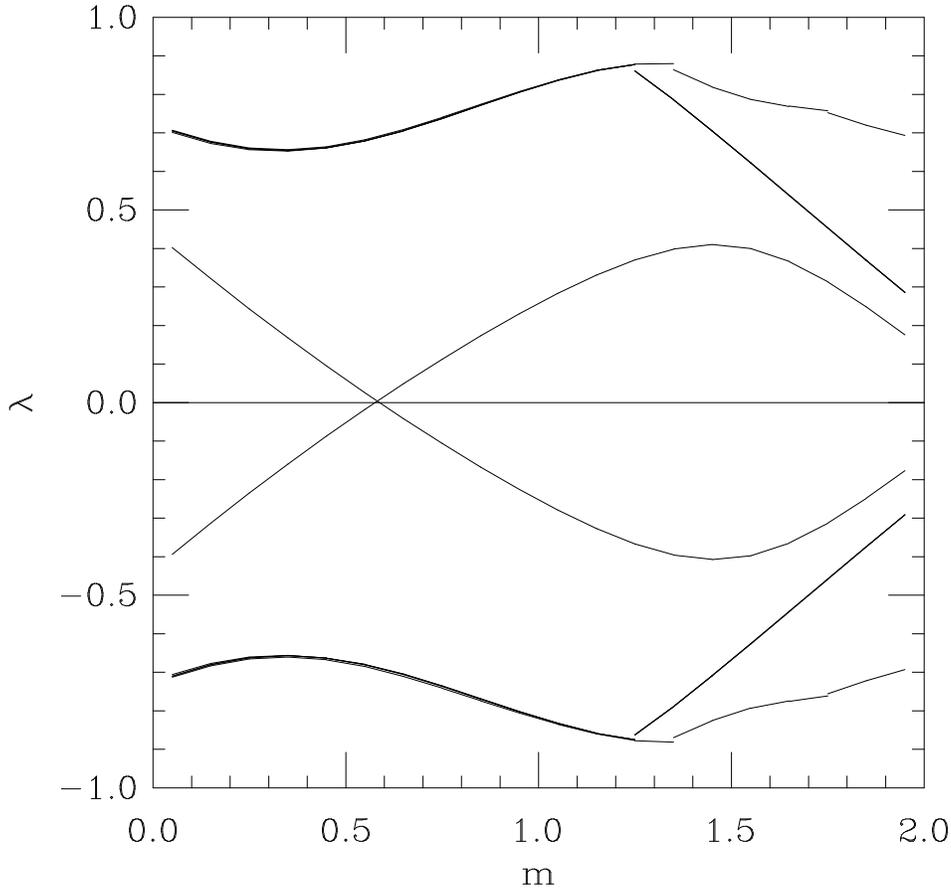}}
\caption{
Spectral flow of $\ham(m)$ for $0\le m \le 2$ for an instanton--anti-instanton
background on an $8^4$ lattice with anti-periodic boundary conditions
for the Wilson fermions. Both instantons have a size of $\rho=1.5$
and are separated by four lattice spacings in every direction. 
}
\label{fig:spec_AI}
\end{figure}
\begin{figure}
\epsfxsize=5.0in
\centerline{\epsffile{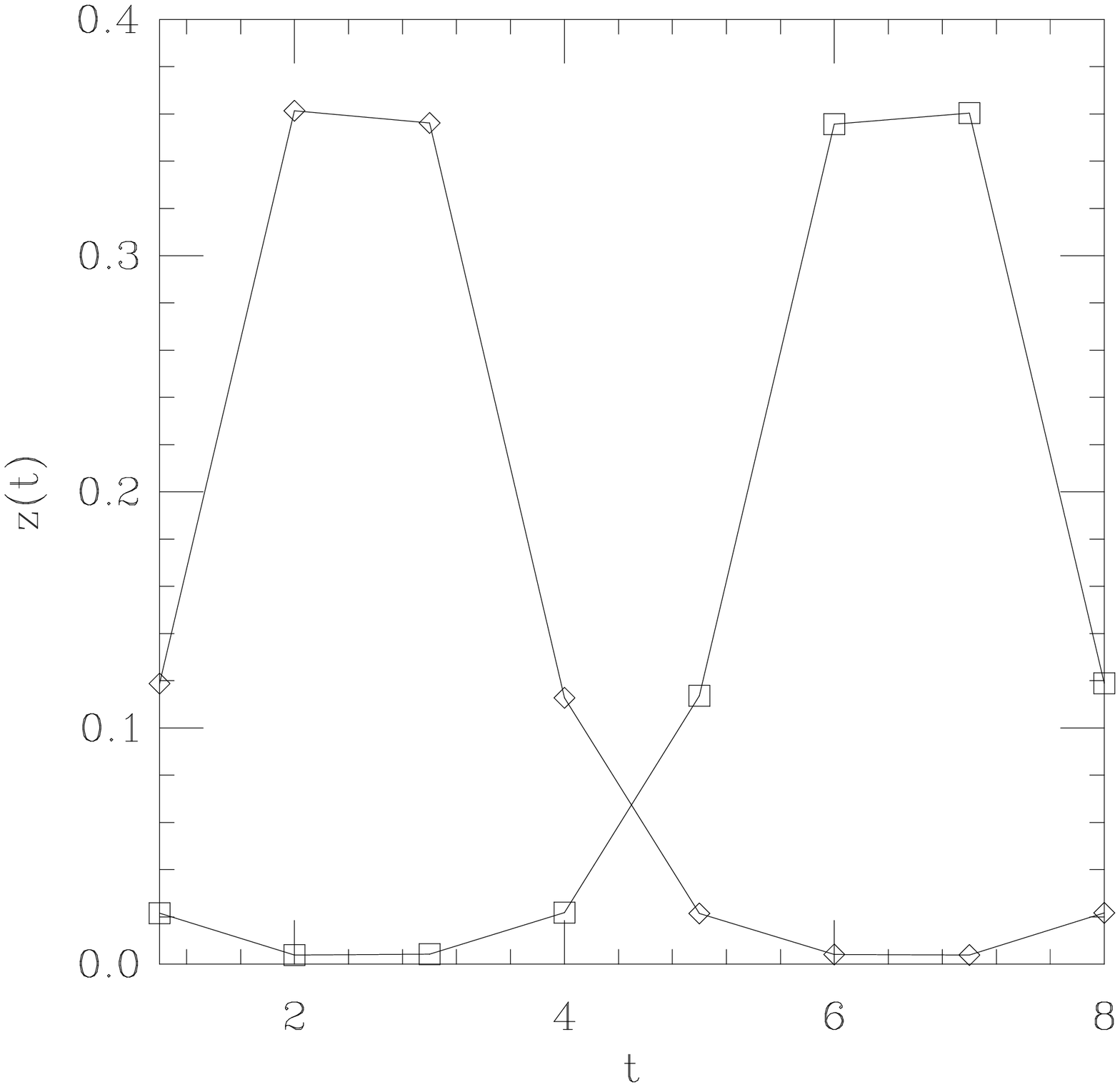}}
\caption{
The modes associated with the level crossings in
Fig.~\ref{fig:spec_AI}.
}
\label{fig:zero_AI}
\end{figure}

\section{Spectral flow in backgrounds with two topological objects}

In the previous section we
established that one can construct a lattice gauge field
configuration close to a continuum instanton
on a finite lattice with toroidal boundary conditions
resulting in a spectral flow of $\ham(m)$ consistent with the
topology.
We now
turn to two instantons and instanton--anti-instanton configurations
as described in section 3.
We set both instantons to have the same size, namely $\rho=1.5$.
We consider the spectral flow of $\ham(m)$ with anti-periodic
boundary conditions on an $8^4$ lattice. We first consider
the situation where the two objects are as far separated as possible
on the $8^4$ lattice, namely the distance between the two objects is
four in all four directions. For the two instanton configuration,
we find that two levels cross downward and are almost degenerate.
The modes associated with these crossings indeed are very close to
the single instanton zero modes with $\rho=1.5$. The spectral flow
is shown in Fig.~\ref{fig:spec_II} and the two modes at the crossing
point are shown in Fig.~\ref{fig:zero_II}.
For the instanton--anti-instanton configuration we find that one level
crosses downwards and another one crosses upwards almost at the same
point in $m$. Like in the two instanton case, these modes are localized
at the site of the instanton and anti-instanton and have a size consistent
with $\rho=1.5$. 
The spectral flow
is shown in Fig.~\ref{fig:spec_AI}, and the two modes at the crossing
point are shown in Fig.~\ref{fig:zero_AI}.
The other extreme is when the two instantons are placed on top of each other.
We found only one level crossing downward indicating that the
configuration had a topological charge of unity. The mode associated with
the crossing is significantly broader than the ones we found for the
well separated situations with the size now consistent with 
$\rho=1.5\times\sqrt{2}$.
If the two instantons were separated by $(2,0,0,0)$, we still found
only one level crossing downward indicating that the configuration still
had a topological charge of unity. The mode associated with the crossing
was narrower than the case where the instantons were on top of each other,
but still broader than the ones found for well separated instantons.
When the instantons were separated by $(4,0,0,0)$, we found two
crossings indicating that the two instantons were sufficiently
separated for them to be resolved as individual instantons.
Similar results were found for instanton--anti-instanton configurations.
When they were on top of each other no level crossings were observed.
When they were separated by a distance of $(2,0,0,0)$ and $(4,0,0,0)$,
we found one level crossing downwards and one level crossing upwards
indicating that the instanton and anti-instanton were resolved for
both these separations.

\section{Conclusions}

We studied the spectral flow of $\ham(m)$ in smooth instanton
backgrounds. For single instanton backgrounds we found the
following:
\begin{itemize}
\item With Dirichlet boundary conditions for $\ham(m)$, we found
a single level crossing in $\ham(m)$ with the eigenmode at the
crossing point describing a proper realization of the continuum zero mode.
\item Instantons with larger sizes resulted in a crossing at smaller
values of $m$, and we found level crossing all the way up to $m=2$
as we reduced $\rho$.
\item Instantons centered in the middle of a lattice hypercube fell
through the lattice if their size was smaller than about one lattice spacing.
The Wilson action for these instantons on an open lattice is a 
monotonic function of $\rho$ with the action going to zero as
$\rho$ goes to zero. The action at $\rho=1$ is well above the
dislocation bound~\cite{Kremer} of $12\pi^2/11$.
\item Instantons centered on a lattice site never fell through the lattice.
This is consistent with the Wilson action for these instantons.
It is not a monotonic function of $\rho$. It has a minimum for a
specific value of $\rho$ ($\rho=1$ on an $8^4$ open lattice) below which
the action increases. The action never goes below the dislocation 
bound~\cite{Kremer}. 
\item With anti-periodic boundary conditions for $\ham(m)$, we found
that instantons in regular gauge resulted in a gauge field configuration
with no level crossing; however, instantons in singular gauge resulted
in a single level crossing consistent with the topology.
\end{itemize}
For two instanton and instanton--anti-instanton backgrounds, we
found level crossings consistent with two topological objects
if the two instantons were separated by a distance of
four or more and if the instanton and anti-instanton were separated by
a distance of two or more. 

\ack{
This research was supported by DOE contracts 
DE-FG05-85ER250000 and DE-FG05-96ER40979.
We would like to thank Khalil Bitar
for useful discussions.
Computations were performed on the CM-2 and the 
workstation cluster at SCRI.}

\end{document}